\documentclass{webofc}
\usepackage[varg]{txfonts}   

\woctitle{Is time enough in order to know where you are?}

\begin{document}

\title{Is time enough in order to know where you are?}

\author{Angelo Tartaglia\inst{1,2}\fnsep\thanks{\email{angelo.tartaglia@polito.it}}
}

\institute{Politecnico di Torino, corso Duca degli Abruzzi 24, 10129 Torino, Italy
\and
INFN, via Pietro Giuria 1, 10126 Torino, Italy}

\abstract{%
  This talk discusses various aspects of the structure of space-time presenting mechanisms leading to the explanation of the "rigidity" of the manifold and to the emergence of time, i.e. of the Lorentzian signature. The proposed ingredient is the analog, in four dimensions, of the deformation energy associated with the common three-dimensional elasticity theory. The inclusion of this additional term in the Lagrangian of empty space-time accounts for gravity as an emergent feature from the microscopic structure of space-time. Once time has legitimately been introduced a global positioning method based on local measurements of proper times between the arrivals of electromagnetic pulses from independent distant sources is presented. The method considers both pulsars as well as artificial emitters located on celestial bodies of the solar system as pulsating beacons to be used for navigation and positioning.
}
\maketitle
\section{Introduction}
\label{intro}
The problem of positioning an event within a four-dimensional manifold, such as space-time, necessarily implies various assumptions concerning what the properties of space-time are. In particular something must be clarified concerning time and its special status with respect to the three space dimensions. For this reason I shall devote the first part of my talk to the emergence of time from the geometrical properties of an "elastically" deformable four-dimensional manifold. The idea of such a space-time is assumed from the Strained State theory developed in ref.s \cite{cqg1}-\cite{nst1}.
Once the origin of the light cones has found a logical framework within which to fit, we may exploit both the geometrical properties of the manifold and the fact that the "length" of a portion of the world-line of any observer is given by the proper time measured by any clock carried along by the traveler. If the measured proper time intervals are between the arrivals of regular pulses emitted by far away sources, whose world-lines are known, I shall show that the time sequences are sufficient to enable the voyager to find out its own way in space-time reconstructing piecewise its world-line and locating it in a previously defined global reference frame. The positioning system developed according to this approach is intrinsically relativistic, in the sense that it automatically includes all relativistic effects with no need for by hand corrections as the ones that are introduced in the GPS or similar systems. The relativistic positioning system based on proper time measurements has been developed and tested in ref.s \cite{acta}-\cite{ref10}. In the following I shall outline the essence of the method.

\section{Space-time and its properties}
\label{sec-1}
\begin{figure*}
\centering
\includegraphics[width=8cm,clip]{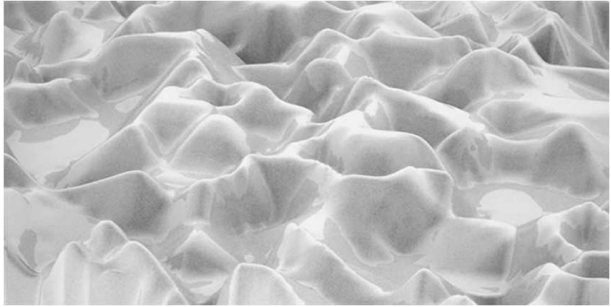}
\caption{Bidimensional representation of a warped space-time. The picture is of course Euclidean, but the space-time would have Lorentzian signature}
\label{fig-1}
\end{figure*}

According to the theory of General Relativity (GR) space-time is a four-dimensional Riemannian manifold with a Lorentzian signature. A pictorial view of a space-time can be seen in fig.~\ref{fig-1}.

The figure is of course bidimensional and the representation has Euclidean signature rather than the true Lorentzian signature of actual space-time, but it gives an idea of the geometrical nature of our manifold, exaggerating its warpedness. The local curvature is a manifestation of the gravitational interaction.

\subsection{Where does time come from?}
\label{sec-2}
In GR curvature is related, through Einstein's equations, to the matter/energy density. When taking all matter/energy away what is left is a flat Minkowski manifold. This simple result sounds rather obvious, because the absence of any source consistently implies a flat manifold. There is however something peculiar to a Minkowski manifold, besides the flatness: it is the presence of the light cones or, otherwise stated, the peculiar signature that singles out one dimension with respect to the other three.
Suppose space-time is not a mere mathematical artifact, but it represents something real. If it was not so, what meaning would we attach to the famous sentence by John Wheeler: "matter tells space-time how to curve, and space-time tells matter how to move"? Space-time will have properties on its own, but any non-trivial geometrical feature will depend on matter/energy. Now, any given manifold with nothing in it should have all possible symmetries and the corresponding geometry. In practice a space-time totally devoid of matter should be flat and have Euclidean geometric properties; consequently it should be perfectly isotropic, which means that no direction could have any properties distinguishing it from the others. Where would the light cones of the Minkowskian geometry come from? But of course we know that the light cones do exist in a real space-time (according to GR, which we assume to hold); they belong to a Minkowski manifold too, as far as the latter is intended as the local tangent space of a generally curved space-time. A local tangent space preserves some symmetry from the original manifold, and in particular the peculiarity of time, i.e. the local light cone; however this is not the case for a space-time totally deprived of matter/energy.

In order to try and find at least a reasonable origin for the special role of time with respect to space, we can explore the possibilities lent by the idea that space-time is indeed a real continuum. The analogy we may exploit is with ordinary three-dimensional continua; we know them to be deformable media, whose behaviour, at least at the lowest order of approximation, is described by the linear elasticity theory. Besides this aspect we also know that material continua can contain such things as structural or texture defects. A defect can be built by the ideal procedure described originally by Volterra \cite{volterra}. Start from a flat four-dimensional Euclidean manifold; imagine to cut out a finite patch of the manifold. By this simple proceeding two distinct families of geodetic lines have been sorted. As far as the manifold is flat, geodesic curves on it are all possible straight lines; cutting out an area we practically distinguish all the complete geodesics (i.e. the ones which do not impinge into the missing region) from the incomplete ones which are limited, on one side, by the border of the prohibited area. This is indeed a first step towards the definition of time-like world-lines contrasted with space-like ones, but does not touch the problem of signature yet.

Adopting the idea that space-time can host defects just in the same sense as ordinary physical continua do, there is one more feature on which we may draw our attention. A defect is not only identifiable with a secluded region of a manifold; we must also think of pulling the rim of the hole inward until the gap is closed and the opposite portions of the border are glued together so that the manifold no longer contains voids but rather "scars". From the geometrical viewpoint the envisaged process is the continuous formation of a singularity in the manifold, where the singularity may be any singular sub-manifold: a point, a line, a bidimensional surface, a hypersurface... What matters now, however, is that the presence of a defect in a material continuum induces a spontaneous strained state, where "spontaneous" means "in the absence of the action of external agents ("forces")". Applying these concepts to space-time amounts to assume that it contains at least one global defect representing the origin of all the geodesics we call time-like and being the cause of curvature even in the absence of matter/energy (the "external agent").

The above thumbnail description is the basis of the cosmic defect theory \cite{cosmo}. It ascribes the global Robertson-Walker (RW) symmetry of the universe to the presence, in the four-dimensional manifold, of a cosmic defect, which is also responsible, via the induced strain, of what we call the accelerated expansion. The strain is accounted for by an additional term in the empty space Einstein-Hilbert Lagrangian density. The additional term is molded on the deformation energy of elastic continua; the basic ingredient is the strain tensor proportional to the difference between the metric tensors of a flat undeformed Euclidean manifold (reference manifold) and of the actual space-time (natural manifold). In terms of the line elements we may write:

\begin{eqnarray}
ds_E^2&=&E_{\mu\nu}dx^{\mu}dx^{\nu} \nonumber \\
\label{linee} \\
ds^2&=&g_{\mu\nu}dx^{\mu}dx^{\nu} \nonumber
\end{eqnarray}

It is understood that the same coordinates are used to identify corresponding events on the two manifolds. The strain tensor is:

\begin{equation}
\sigma_{\mu\nu}=\frac{1}{2}(g_{\mu\nu}-E_{\mu\nu})
\label{strain}
\end{equation}

According to the approach I am presenting here, the full action for the strained state of space-time is \cite{cosmo}:

\begin{equation}
S = \int{(R+\frac{\lambda}{2}\epsilon^2+\mu \epsilon_{\alpha\beta}\epsilon^{\alpha\beta})\sqrt{-g}d^4x}
\label{azione}
\end{equation}

$R$ is the scalar curvature; $\epsilon = \epsilon_{\alpha}^{\alpha}$ is the trace of the strain tensor; $\lambda$ and $\mu$ are the Lam\'{e} coefficients of space-time with exactly the same role as in three-dimensional elasticity. Indices are lowered and raised by means of the full metric tensor $g_{\mu\nu}$ and its inverse. The integrand is the Lagrangian density $L$.

\subsection{A Robertson-Walker space-time}
\label{sec-3}

Let us apply the approach described in the previous section to a Robertson-Walker space-time, i.e. to a manifold homogeneous and isotropic in space. This is the symmetry we think we see in the universe, but here matter is not included for the moment.

Equations (\ref{linee}) may now be written as:

\begin{eqnarray}
ds_E^2&=&b^2(\tau)\tau^2 + dx^2 + dy^2 + dz^2 \nonumber \\
\label{RW} \\
ds^2&=&d\tau^2-a^2(\tau)(dx^2+dy^2+dz^2) \nonumber
\end{eqnarray}

\begin{figure*}
\centering
\includegraphics[width=8cm,clip]{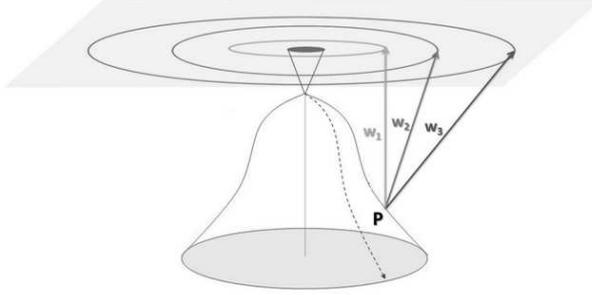}
\caption{Schematic view of a Robertson-Walker space-time embedded in a flat three-dimensional manifold; in this example the space is finite (like in a closed universe). The arrows point out different deformation strategies leading from the flat Euclidean reference manifold to the final natural manifold. The singular vertex of the natural space-time is a defect in as much as it corresponds to a full circular area in the reference manifold. The dotted line is a typical incomplete time-like geodesic that emerges from the defect.}
\label{fig-2}
\end{figure*}

We recognize the typical scale factor $a(\tau)$ depending on the cosmic time $\tau$ expressed as a length. The $b(\tau)$ of the reference manifold does not change the flatness of the Euclidean manifold, but it represents a "gauge" function expressing the fact that there are in principle infinitely many different deformation strategies leading from the reference to the natural manifold, all preserving the global symmetry. The meaning of $b$ is easily understood looking at fig. \ref{fig-2} \cite{cina}.

From eq.s (\ref{RW}) the Robertson-Walker strain tensor immediately follows:

\begin{eqnarray}
\epsilon_{00}&=& \frac{1-b^2}{2} \nonumber \\
\label{RWstrain} \\
\epsilon_{xx}&=& \epsilon_{yy}=\epsilon_{zz} = -\frac{1+a^2}{2} \nonumber
\end{eqnarray}

Once we have the strain tensor we can write the explicit form of the action integral (\ref{azione}) and from it deduce the Euler-Lagrange equations for the unknown functions. The first step is quite simple and leads us to the gauge function $b$ that follows from $\partial L/\partial b = 0$; it is \cite{nst1}:

\begin{equation}
b^2 = 2\frac{2\lambda+\mu}{\lambda+2\mu}+\frac{3}{a^2}\frac{\lambda}{\lambda+2\mu}
\label{b}
\end{equation}

Consequently the equation for the scale factor $a(\tau)$ is:

\begin{equation}
2(2a\ddot{a}+\dot{a}^2)-\frac{\mu}{2a^2}\frac{2\lambda+\mu}{\lambda+2\mu}(3a^4+2a^2-1)=0
\label{a}
\end{equation}

Using the energy condition it is possible to pass to a first order differential equation:

\begin{equation}
W=6a\dot{a}^2-\frac{3\mu}{2a}\frac{2\lambda+\mu}{\lambda+2\mu}(1+a^2)^2
\label{energy}
\end{equation}

From eq. (\ref{energy}) it is possible to desume the square of the Hubble parameter:

\begin{equation}
H^2=\frac{\dot{a}^2}{a^2}=\frac{1}{6a^3}(W+\frac{3}{2}\mu B\frac{(1+a^2)^2}{a})
\label{hub2}
\end{equation}
where the shorthand notation
\begin{equation}
B=\frac{2\lambda+\mu}{\lambda+2\mu}
\nonumber
\end{equation}
has been used.

The next step is to consider that letting $\mu\rightarrow 0$ should bring about the traditional GR result (no strain contribution), which implies $W=0$. Finally we have:

\begin{equation}
\frac{\dot{a}}{a}=\pm\sqrt{\frac{\mu}{4}B}\frac{1+a^2}{a^2}
\label{hubble}
\end{equation}

The double sign in front of eq. (\ref{hubble}) expresses two options: an expanding or a contracting space-time. Since we know the universe is expanding, we choose the $+$ sign.

I stress the fact that this solution has been obtained even in the absence of matter under the assumption of the RW symmetry and of space-time being an "elastic" manifold. These assumptions imply the presence of an initial singularity in the form of a texture defect of the manifold.

\subsection{Signature flip and the emergent rigidity of the manifold}
\label{sec-4}
I had left open the question about the origin of the signature of our space-time. Now considering eq. (\ref{hubble}) we can solve it for $a$, finding \cite{nst1}:

\begin{equation}
a^2=C\exp{\sqrt{\mu B}\tau}-1
\label{segna}
\end{equation}

$C$ is an integration constant. If it is smaller than $1$, we see that, close to the origin of the $\tau$ coordinate, $a$ acquires imaginary values ($a^2<0$). The interpretation of this result is that the manifold has a defect for $\tau=0$; the defect is surrounded by a curved region with Euclidean signature, whose boundary is at
\begin{equation}
\tau=\tau_h=\frac{1}{\sqrt{\mu B}}\ln{\frac{1}{C}}.
\nonumber
\end{equation}

The global RW symmetry is preserved. Below $\tau=\tau_h$ three out of four \emph{space} dimensions are homogeneous and isotropic, but the fourth is not a time at all: there are no light cones. In the Euclidean signature region $\tau$ is just a running coordinate along the space-like incomplete geodesics, that start at the defect in $\tau=0$. On our side of the singular $\tau=\tau_h$ hypersurface we find a Lorentzian signature so that now $\tau$, which still is a running coordinate along the incomplete geodesics stemming out of the defect, acquires a time character and indeed is read as the cosmic time. In the Euclidean signature domain the three homogeneous space dimensions shrink with increasing distance from the defect (increasing $\tau$); in the Lorentzian signature domain they expand.

The theory that I have sketched until now has been tested against the observation of the universe at the cosmic scale and the results have been good \cite{nst2} \cite{nst3}. Remarkably the best fit values of the Lam\'{e} coefficients are $\sim 10^{-52}$ m$^{-2}$; such small values imply that space-time behaves like an extremely rigid stuff. On the other side the whole description so far is based on an analogy with the three-dimensional elasticity theory and we know that that theory emerges from the properties of a microscopic discrete structure underlying the apparently continuous aspect of solid materials. We are then led to think that this could also be the case of space-time.

\begin{figure*}
\centering
\includegraphics[width=8cm,clip]{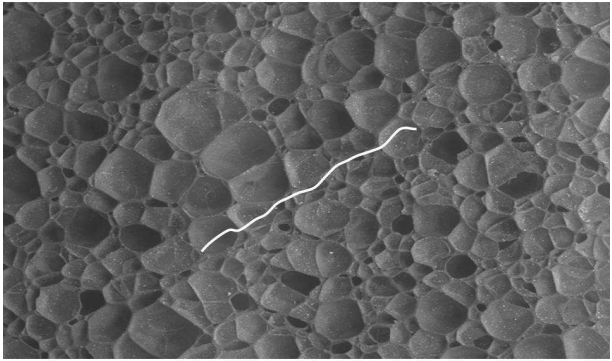}
\caption{A geodetic line in a foamy manifold. At a higher scale the line would appear to be straight.}
\label{fig-3}
\end{figure*}

Following this possibility we may for instance remark that high rigidities even of \emph{per se} soft materials are attained when the material has a foamy structure. Why then not consider that space-time too has, at microscopic level (Plank scale?), the topology of a four-dimensional foam? This would produce an extremely rigid behavior at macroscopic as well as cosmic scale. A typical geodetic line would then appear, at the microscopic level, as in fig. \ref{fig-3}. At a higher scale the geodetic would practically be a straight line.

I would like to stress that the idea of an underlying foamy topology is here entirely classical. I am not calling in any specific attempt to quantize gravity. The whole conceptual framework in which the strained state theory is cast is classical or, maybe, effectively classical, and it is applied essentially at the cosmic or at least astronomical scale. Given the numerical values of the Lam\'{e} coefficients, no relevant effect is expected at the local (for example at the solar system) scale.

\section{Positioning}
\label{sec-5}

After having given a logical frame for the presence and relevance of time, I may try and answer to the question in the title of the present article coming to the practical problem of positioning, i.e. of finding the position of an observer within space-time.

Any attempt to set up a global positioning system has a number of underlying assumptions which I am recalling here:
\begin{itemize}
  \item  General Relativity holds, i.e. space and time are tied to each other by geometrical laws;
  \item  (our) space-time is a four-dimensional Riemannian manifold with Lorentzian signature;
  \item  it is possible to set up a global reference frame within which local coordinates frames can be defined.
\end{itemize}

\begin{figure*}
\centering
\includegraphics[height = 60 mm, width=130 mm]{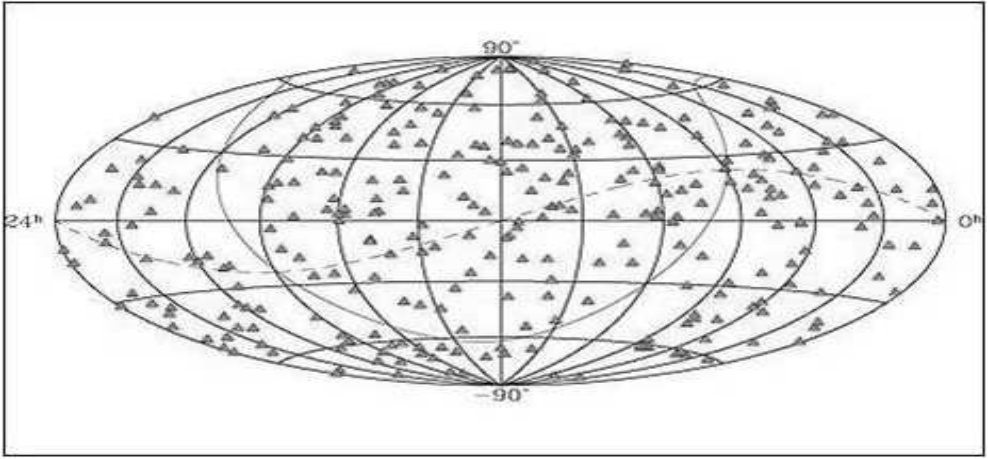}
\caption{The distribution in the sky of a few quasars taken from the Sloan Digital Survey}\label{fig-4}
\end{figure*}

The first two assumptions were implicit in all I have written in the previous sections. As for the global reference frame, currently it is assumed to be attached to the "fixed stars" intended as quasars. Quasars are indeed assumed to be at distances in the order of billions of light years so that their reciprocal positions in the sky may be treated as fixed, notwithstanding their proper motions and for times as long as the human history. Fig. \ref{fig-4} gives an example of the positions of some quasars in the sky. They identify a corresponding bunch of fixed directions that are the same for all observers in the solar system with accuracies better than $10^{-12}$.

Actually the use of quasars implies one more hypothesis which, strictly speaking, is improper but may be assumed to be approximately or effectively true:
\begin{itemize}
\item space-time is asymptotically flat,
\end{itemize}
i.e. quasars, as point-like objects (!), are represented by straight parallel world-lines.

\subsection{Null geodesics}
\label{sec-6}
\begin{figure*}[]
\centering
\includegraphics[height = 78 mm, width=130 mm]{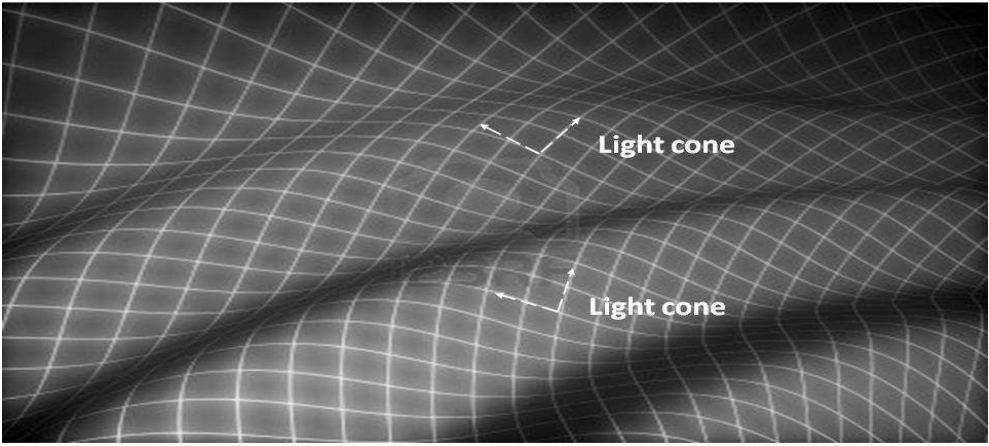}
\caption{Two sets of null geodesics covering a bidimensional curved manifold}\label{fig-5}
\end{figure*}

Once these assumptions have been accepted, a good strategy in order to define a set of coordinates is to rely on null geodesics. Four independent families of such geodesics covering all space-time are a good means to identify any event there as the unique intersection among four geodesics each from a different family. The bidimensional sketch in fig. \ref{fig-5} gives the idea.

Each null geodesic is locally identified by its null tangent four-vector, written as:

\begin{equation}
\chi=cT(1,\cos{\alpha},\cos{\beta},\cos{\gamma})=cT(1,\hat{n})
\label{tangent}
\end{equation}

It is $\chi^2=0$. The space components of (\ref{tangent}) are the direction cosines of the line with respect to the local axes of the reference frame. The factor in front can have any value without modifying the geometrical meaning of $\chi$; it can be used to host some additional information: here it contains $T$ which is the period of the electromagnetic signal propagating along the geodesic.

If the space-time is flat (\ref{tangent}) identifies not only a specific geodesic of a given family, but the whole family everywhere.

Four independent vectors like (\ref{tangent}) form a null basis, that can be used to represent any four-vector $r$, pointing to any position in the surroundings of an origin. It will be:

\begin{equation}
r=\frac{\tau_1}{T_1}\chi_1+\frac{\tau_2}{T_2}\chi_2+\frac{\tau_3}{T_3}\chi_3+\frac{\tau_4}{T_4}\chi_4
\label{vettore}
\end{equation}

The pure numbers $\tau_a/T_a$ (Latin letters from the first part of the alphabet label the families of geodesics: $a=1,2,3,4$) are called light coordinates of the event on the tip of the vector.

A complementary view to the one based on null tangent vectors is based on the hypersurfaces dual to them:

\begin{equation}
\varpi^{abc}=\epsilon^{abcd}\chi_d
\label{hyper}
\end{equation}

\begin{figure*}
\centering
\includegraphics[height = 70 mm, width=90 mm]{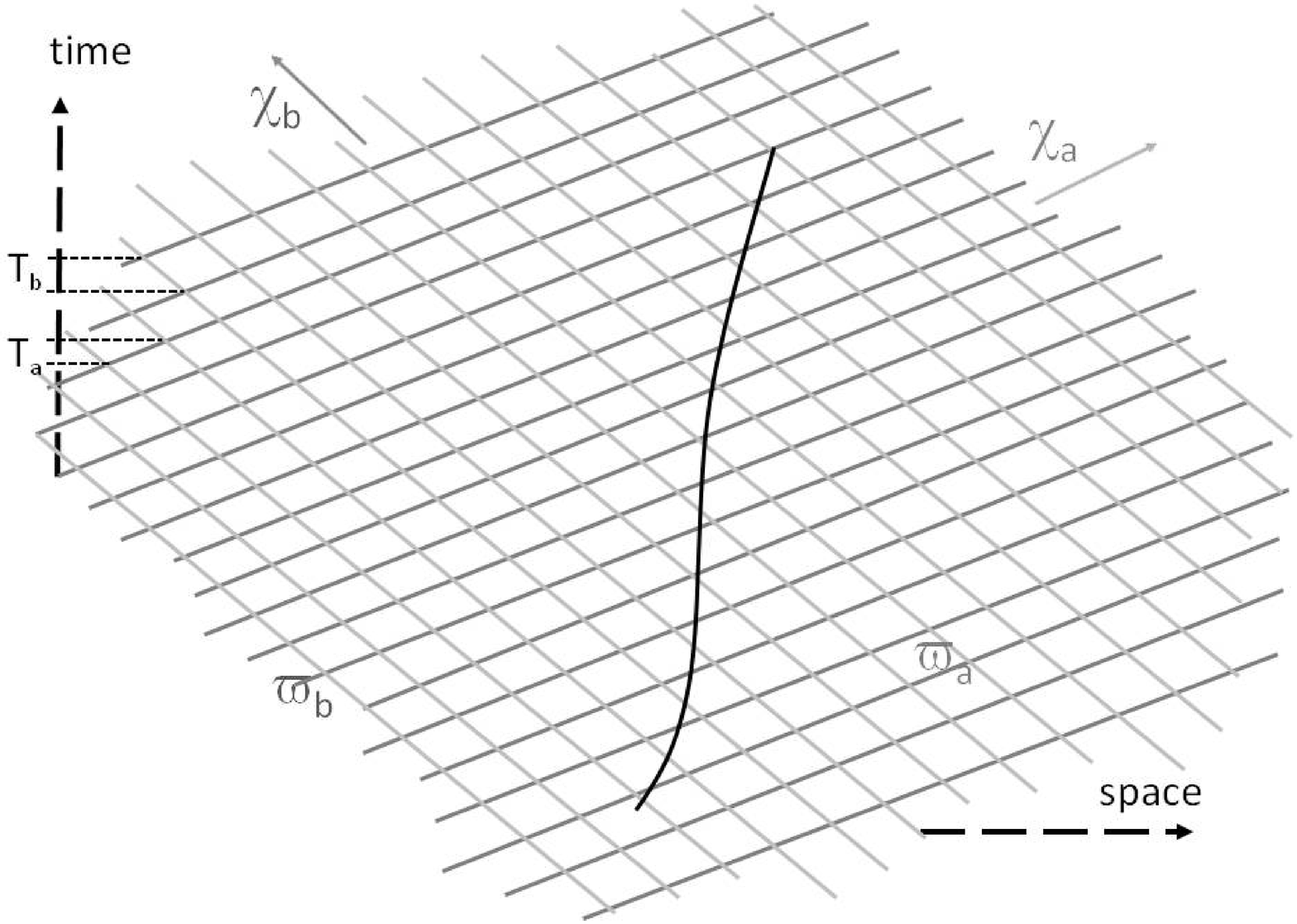}
\caption{A bidimensional flat space-time covered by a grid made of null hypersurfaces (actually lines) conjugated to the null vectors $\chi_{a,b}$. The wavy line is the world-line of an observer.}\label{fig-6}
\end{figure*}

$\epsilon_{abcd}$ is the fully antisymmetric Levi-Civita tensor. If the space-time is flat then the $\varpi^{abc}$'s represent four families of null hyperplanes, covering the whole manifold. Otherwise $\varpi^{abc}$ identifies the local tangent space to one of the null hypersurfaces perpendicular to the corresponding $\chi$. Hypersurfaces from the same family never intersect each other so that four from different families intersect at only one event in the manifold, thus uniquely identifying its position. The situation is represented in two dimensions in fig. \ref{fig-6}. Here local singularities and horizons are not taken into account.

A grid is shown built by hyperplanes spaced out by the period of the signal from each source. Any world-line can in principle be identified by the intersections of the hypersurfaces of the grid.

\subsection{Finding the light coordinates}
\label{sec-7}
A practical implementation of the principles stated in the previous section may be obtained using discrete electromagnetic pulses coming from (not less than) four independent sources located at infinity; the $a$-th source emits  pulses at the rate of $1/T_a$ per second. The $T$ parameter of formula (\ref{tangent}) is now interpreted as the repetition time of the pulses rather than the period of a monochromatic continuous wave. The grid exemplified in fig. \ref{fig-6} is now really discrete; we have then a sort of an egg-crate whose walls are in a sense "thick" because they are associated to pulses which have, though short, a duration in time: in practice the hypersurfaces on the graph correspond to "sandwich waves" carrying a pulse. The sides of the cells are measured by the $T$'s, projected along the time axis of the background global reference frame (we should remember that the bundles of hypersurfaces forming the walls of the cells are all null).

\begin{figure*}
\centering
\includegraphics[height = 80 mm, width=110 mm]{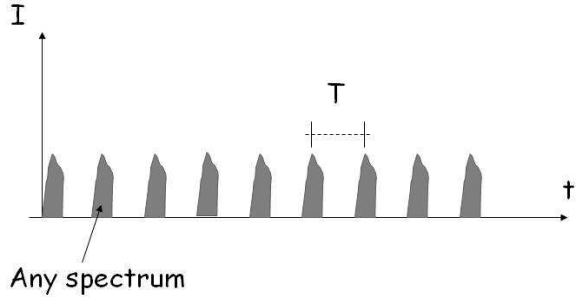}
\caption{Typical emission sequence of the pulses from a source. Vertically intensities are drawn; the profile of the pulse is not important; times are proper times of the emitter.}\label{fig-7}
\end{figure*}

The world-line of an observer necessarily crosses the walls of successive boxes of the egg crate. If we are able to label each cell of the crate assigning integer numbers to the walls, we are also able to reconstruct the position of the observer in the manifold.  A typical emission diagram of one of the sources will more or less be like the one sketched in fig. \ref{fig-7}.

The shape of the pulse is not important as well as it is not the spectral content of it. What matters is its reproducibility and the stability of the repetition time. Considering natural pulses, as the ones coming from pulsars, we find repetition times ranging from several seconds down to a few milliseconds and lasting a fraction of the period. As an example of artificial pulses the highest performance is obtained with lasers: GHz frequencies are possible with pulses as short as $\sim10^{-15}$ s.

\begin{figure*}
\centering
\includegraphics[height = 80 mm, width=110 mm]{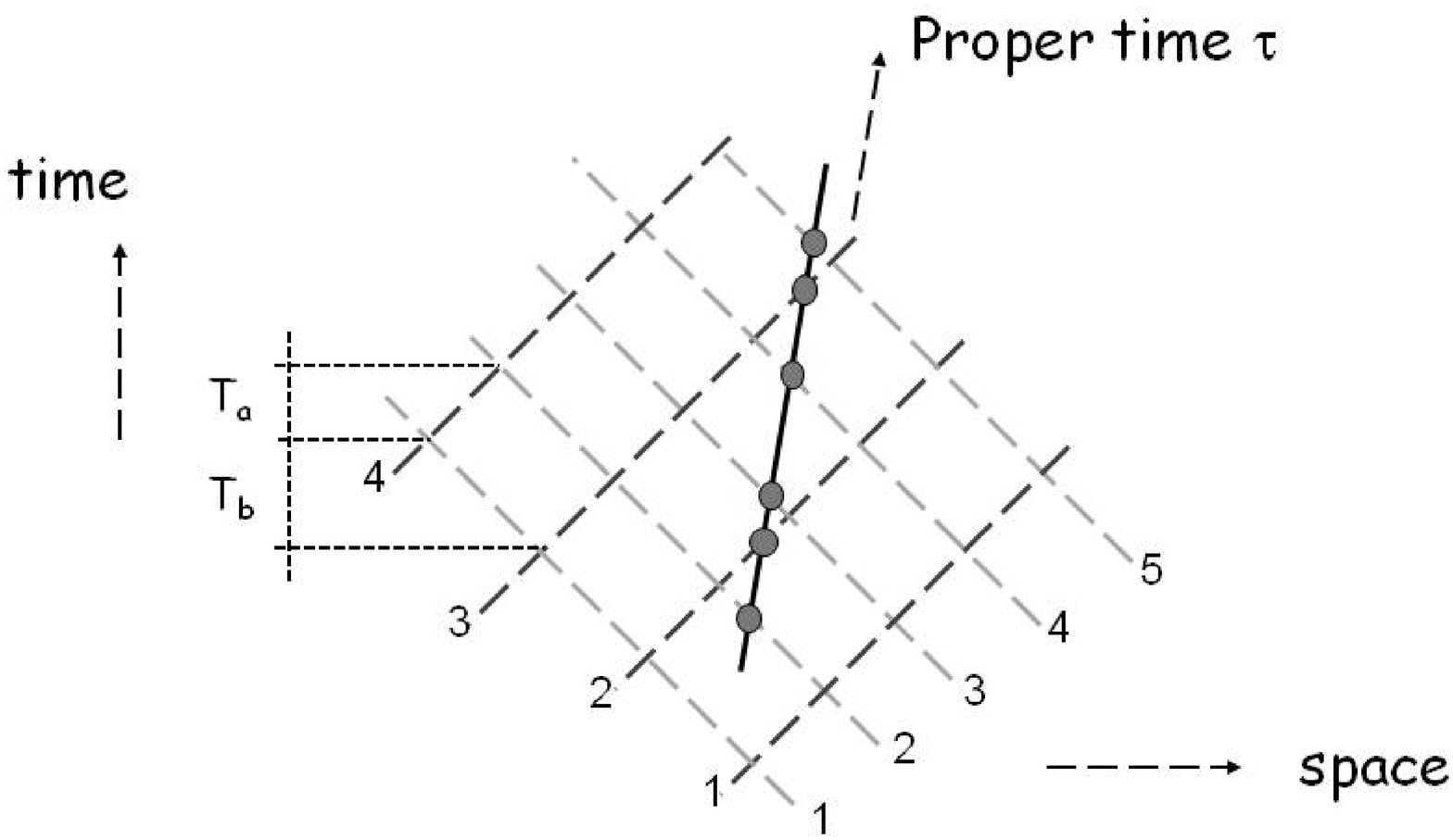
}
\caption{A straight portion of a world-line is shown in a local flat patch of space-time. The lines of the grid correspond to different pulses labeled by their ordinal integer. The intersections of the world-line with the walls are localized by quadruples (actually pairs in the figure) of real numbers one of which is always an integer.}\label{fig-8}
\end{figure*}

Once pulses are used, we may label them in order, by integer numbers, as it can schematically be seen in fig. \ref{fig-8}.

The integers can be though of as rough coordinates identifying the cells of the grid. At this level the approximation would be rather poor, being of the order of the size of each cell. If the periods are milliseconds this corresponds to hundreds of kilometers. Looking at fig. \ref{fig-8} we may however notice that the intersections of a given world-line with the walls of the cells are labeled by a quadruple of numbers, at least one of which is an integer: these numbers are the coordinates of the intersection points. We may write the typical light coordinate of a position in the crate as
\begin{equation}
\frac{\tau}{T}=n+x
\label{coordi}
\end{equation}
The $n$'s are the integers, whilst the $x$'s are the fractional parts.
If we have a means to determine the $x$'s the localization of an intersection event can be done with an accuracy much better than the hundreds of km I mentioned above.

Considering that the intersections coincide with the arrivals of pulses from different sources, the determination of the fractional part of the coordinates is indeed a trivial task, provided the traveler carries a clock, the space-time is flat and the world-line is straight. Once one measures the proper intervals between the arrivals of successive pulses, a simple linear algorithm based on elementary four-dimensional flat geometry produces the $x$'s \cite{ref11}. In fact the proper time interval between the arrivals of the $i$-th and the $j$-th pulses from a given source is the norm of the $r_{ij}$ four-vector separating the two arrival events, i.e.
\begin{equation}
\tau_{ij}=|r_j-r_i|=|(X_{aj}-X_{ai})\chi^a|=|\Delta X_{aij}\chi^a| \nonumber
\end{equation}

Simple proportions tell us that \cite{ref11}:
\begin{equation}
\frac{\tau_{ij}}{\tau_{jk}}=\frac{\Delta X_{1ij}}{\Delta X_{1jk}}=\frac{\Delta X_{2ij}}{\Delta X_{2jk}}=\frac{\Delta X_{3ij}}{\Delta X_{3jk}}=\frac{\Delta X_{4ij}}{\Delta X_{4jk}}
\label{prop}
\end{equation}
and eight successive arrival events are sufficient for determining all the $x$'s of the sequence.
It is:

\begin{eqnarray}
x_{a1}&=&0, \hspace{0.4cm} x_{b1}=1-\frac{\tau_{12}}{\tau_{26}},  x_{c1}=1-\frac{\tau_{13}}{\tau_{37}}, \hspace{0.8cm} x_{d1}=1-\frac{\tau_{14}}{\tau_{48}} \nonumber \\
x_{a2}&=&\frac{\tau_{12}}{\tau_{1,5}},  x_{b2}=1, \hspace{0.8cm} x_{c2}=1-\frac{\tau_{13}}{\tau_{37}}+\frac{\tau_{12}}{\tau_{37}}, x_{d2}=1-\frac{\tau_{14}}{\tau_{48}}+\frac{\tau_{12}}{\tau_{48}} \label{coo} \\
x_{a3}&=&.......... \nonumber
\end{eqnarray}

Using moving sets of eight successive arrivals we piecewise reconstruct the whole world-line of the receiver. The accuracy of the result depends on the precision of the clock which is being used in order to measure the proper intervals between pulses and on the stability of the period of the pulses, which in turn tells us what the effective "thickness" of the walls of the cells of our space-time crate is. Just to fix some order of magnitude, let me remark that nowadays to have a portable clock with a $10^{-10}$s accuracy is quite easy (much better performances can be achieved in the lab); on the other side, considering pulsars, we have some, whose period is known and stable down to $10^{-15}$s. With these figures the final positioning can be within a few centimeters.

Of course the traveler's motion will not in general be an inertial one and space-time will not be flat, however a short enough stretch of the world-line can always be confused with the tangent straight line to it and a small enough patch of space-time can always be confused with a portion of the local tangent space. In practice we work on the local tangent space and on a linearized portion of the world-line. The acceptability of these assumptions depends on the accuracy required for the positioning and on the constraints posed by the linear algorithm in use. If $\delta \tau$ is the maximum proper time inaccuracy that we decide to be tolerable, the final relative accuracy of the positioning will be \cite{ref11}:

\begin{equation}
\label{delta}
|\frac{\delta x}{x}|\leq 4(\frac{1}{\tau_{i,i+4n}}+\frac{\tau_{i,i+1}}{\tau_{i,i+4n}^2})\delta\tau
\end{equation}

The index $i$ in eq. (\ref{delta}) labels the order of the arrival events; $\tau_{i,i+4n}$ is the proper time interval between the $i$-th and the $(i+4n)$-th arrival, being $n\geq1$ an integer; $n$ should assume the highest value compatible with the straightness hypothesis for the world-line. Of course the number of pitches that can safely be considered depends on the periods $T_a$ of the emitting sources: the shorter are the periods, the bigger is the number of paces that can be used within the linearity assumption.

A pictorial view of what we are doing is as follows. Imagine to embed the real four-dimensional manifold, together with its tangent space at the start event, in a five-dimensional flat manifold; then consider the real world-line of the traveler and project it onto the tangent space. The world-line on the tangent space is what we are piecewise reconstructing by our linear algorithm: in practice we are building a flat chart containing the projection of our space-time trajectory. The time dependence of the adimensional coordinates of the projected world-line may of course be written in the form of a power series, as:

\begin{equation}
\label{serie}
n_a+x_a=u_a\frac{\tau}{T_a}+\frac{1}{2}\alpha_a\frac{\tau^2}{T_a^2}+...
\end{equation}

The coefficients $u_a$ and $\alpha_a$ are proportional to the four-velocity and four-acceleration of the traveler. The individual segments used for the reconstruction are short enough so that the second and further terms of (\ref{serie}) are negligible with respect to the linear one. Given the tolerance $\delta\tau$ on the time measurement, the maximum acceptable duration of an elementary sequence will be:

\begin{equation}
\label{max}
\tau_{max}=\sqrt{2|\frac{u_a}{\alpha_a}|\delta\tau}
\end{equation}

Going on, after a number of paces, the possible presence of an extrinsic curvature of the projected world-line shows up; we know that locally it is impossible to distinguish a gravitational field from a non-gravitational acceleration so we need additional information for that purpose. In the case of a gravitational field evidenced by the reconstruction process I am describing, we get from the data the gradient of the Newtonian gravitational potential $\Phi$. Actually it becomes visible when

\begin{equation}
\label{newt}
|\hat{u}\cdot\overrightarrow{\nabla}\Phi|\geq 4\frac{\delta\tau}{\tau^2}
\end{equation}

In order not to cumulate the distortion introduced by the projection from the real curved manifold to the tangent space at a given event, we need periodically to restart from a further event on the world-line, i.e. to pass to the tangent space at a different event. If the visible curvature of the line on the tangent space as well as the tilt of the successive tangent spaces continues for long in the same sense, the linearization process, as in all similar cases, tends to produce a growing systematic discrepancy with respect to the real world-line, so that periodically one has to have recourse to some independent position fixing method in order to reset the procedure.

\subsection{Possible sources}
\label{sec-8}
\subsubsection{Pulsars}
\label{sec-9}
Possible natural sources of pulses are pulsars. This kind of neutron stars are indeed good pulse emitters because of their extreme stability and long duration. As we know, their emission is in the form of a continuous beam. The apparent periodicity is due to the fact that the emission axis (the magnetic axis) does not coincide with the spin axis of the object so that it steadily rotates, together with the whole star, about the direction of the angular momentum. The pulses arise from the periodic illumination of the earth by the rotating beam. The stability is guaranteed by the angular momentum conservation.

The idea of using pulsars for positioning and navigation was put forth almost as soon as they were discovered \cite{downs} and indeed the advantages of this kind of sources are numerous. Their period is extremely stable and is sometimes known with the accuracy of $10^{-15}$ s; it tends to decay slowly (the relevant times are at least months), but with a very well known trend, determined by the emission of gravitational radiation. Typically the fractional decay rate of the period is in the order of one part in $10^{12}$ per year. The number of such sources is rather high, so that redundancy in the choice of the sources is not a problem: at present approximately 2000 pulsars are known and their number continues to increase year after year. Being these stars at distances of thousands of light years from the earth, they can be treated as being practically fixed in the sky; in any case their slow apparent motion in the sky is known, so corrections for the position are easily introduced.

Unfortunately pulsars have also major drawbacks. One is that their distribution in the sky is uneven, since they are mostly concentrated in the galactic plane, which fact brings about the so called "geometric dilution" of the accuracy of the final positioning: sources located on the same side of the observer produce an amplification of the inaccuracy originating in the intrinsic uncertainties. Furthermore individual pulses differ in shape from one another so that some integration time is needed in order to reconstruct a fiducial series of pulses; this fact, also considering the length of the repetition time, can conflict with the linearization of the world-line of the traveler. It should also be mentioned that most pulsars are subject to sudden jumps in the frequency (glitches), caused by matter falling onto the star; these unpredictable changes can be made unoffensive by means of redundancy, i.e. making use of more than four sources at a time.

However the most relevant inconvenience with pulsars is their extreme faintness. In the radio domain their signals can be even 50 dB below the noise at the corresponding frequencies; to overcome this problem big antennas are required and convenient integration times accompanied with "folding" techniques must be employed. In principle at least four different sources must be looked at simultaneously and this is not an easy task, especially with huge antennas.
The weakness problem has led to consider X-ray- rather than radio-pulsars for positioning \cite{sheikh}. A few hundreds X-ray emitting pulsars are indeed known; their signals are weak too, and can be received only outside the atmosphere, but the background noise is far smaller than the one typical in the radio domain; as for the hardware, X-ray antennas can be much smaller than the typical radio-antennas. 

\begin{figure*}
\centering
\includegraphics[height = 80 mm, width=110 mm]{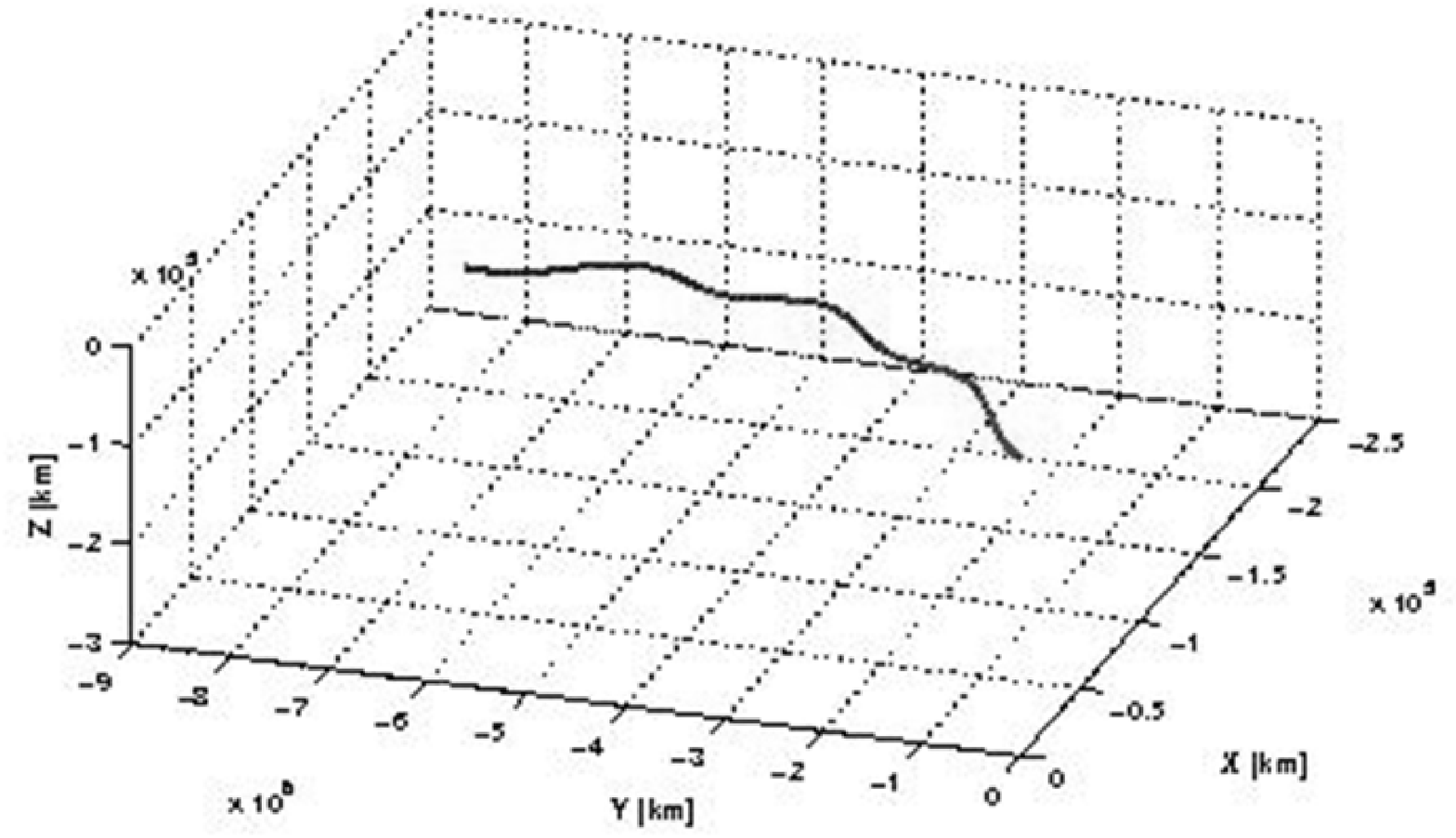}
\caption{The picture shows the motion of the Parkes observatory carried by the earth in its rotation and revolution motion, during three days. The reconstruction has been made applying the relativistic positioning method to the simulated arrival times of the signals from four real pulsars. The reconstructed trajectory is superposed to the real one.}\label{fig-9}
\end{figure*}

The principle feasibility of a pulsar based positioning system, applying the method I have described in the previous sections, has been tested by a simulated exercise named "Eppur si muove" \cite{ref10}. Using a software employed by astrophysicists in order to forecast the arrival times of the pulses from known pulsars at any point of the earth (the name of the software is TEMPO2), we have mimicked an antenna located at the Parkes observatory in Australia and the pulses from four real pulsars there, during three days. The method has been able to reconstruct the motion of the chosen location, together with the whole earth, with respect to the fixed stars, represented in this case by the pulsars. Fig. \ref{fig-9} shows the result, evidencing the wiggling motion of the Parkes observatory due to the combination of the revolution of our planet around the sun with the diurnal rotation.

\subsubsection{Artificial emitters}
\label{sec-10}
In principle what can be done using pulsars can as well be done by means of artificial emitters of electromagnetic pulses. Artificial emitters can have far higher intensities than pulsars; the repetition time can easily be in the range of $ns$ or less, thus making the linearization process more reliable. The stability of the source over time is not as good as for pulsars, but this can represent no inconvenience as far as the number of sources is redundant and they are kept under control. A problem is in the sources clearly not being at infinite distance, which implies a more complicated geometry and of course the need for a good knowledge of the world-line of the emitter in the background reference frame.

One could think of building a Solar System reference frame made of pulse emitters laid down on the surface of various celestial bodies whose orbits are well known and reproducible: the earth of course, the moon, Mars, maybe some of the asteroids; even some space station following a well defined, highly stable orbit around the sun or a planet.

A blended solution for self-guided navigation in the solar system could combine some artificial emitters, as quoted above, together with a limited number of pulsars (the most intensely emitting ones).

\section{Conclusion}
\label{sec-6}
As we have seen, the answer to the question posed in the beginning is \textit{yes}: it is indeed possible to determine the position of a given event, with respect to a predefined reference frame, just measuring the time sequences of the arrivals of electromagnetic pulses from at least four emitters whose world-line is known. After clarifying the special role of time as stemming from the strain induced in space-time by the presence of a global defect responsible for the Robertson-Walker symmetry of the universe at high enough scales, I have expounded the conditions under which and the method whereby the self positioning is possible. The approach is indeed intrinsically relativistic from the beginning since it relies on the very structure of space-time in order to reconstruct the world-line of an observer. This principle feature, together with the extreme accuracy with which technology allows for very precise measurements of proper times, makes  the proposed RPS (Relativistic Positioning System) rather appealing for Global Positioning purposes and especially for navigation across the solar system, where other methods in use are either impracticable or inaccurate. We may then legitimately expect that the new relativistic method will be implemented in the next generations of global positioning systems. The process will however probably be rather slow because of the pervasive presence of more traditional systems like GPS, which, though less satisfying from the principle viewpoint, have an enormous accumulated advantage due to the huge investments made to implement them for military reasons.

Notwithstanding the expected slow implementation for the most practical aims in the terrestrial environment, the RPS can also be the basis of space-time geodesy for fundamental physics objectives. Consider for instance a swarm of satellites orbiting the earth and allow them to exchange with each other electromagnetic pulses (for instance laser pulses): accurate timing of the travel times of the pulses within the swarm would permit to map the region of space-time where the satellites are located, evidencing the average curvature, i.e. the gravitational field. For long enough base lines even propagating disturbances of the curvature (i.e. gravitational waves) could be detected. So finally let me conclude that light is indeed an excellent probe of the structure of space-time especially when coupled with accurate local measurements of proper time intervals.

\end{document}